\documentstyle[amssymb,preprint,prd,aps]{revtex}

\global\arraycolsep=2pt
\makeatletter
\def\fmslash{\@ifnextchar[{\fmsl@sh}{\fmsl@sh[0mu]}}
\def\fmsl@sh[#1]#2{  \mathchoice
    {\@fmsl@sh\displaystyle{#1}{#2}}    {\@fmsl@sh\textstyle{#1}{#2}}    
{\@fmsl@sh\scriptstyle{#1}{#2}}    {\@fmsl@sh\scriptscriptstyle{#1}{#2}}}
\def\@fmsl@sh#1#2#3{\m@th\ooalign{$\hfil#1\mkern#2/\hfil$\crcr$#1#3$}}
\makeatother

\begin{document}
\draft
\title{A Necessary And Sufficient Condition of Distillability with unite fidelity
from Finite Copies of a Mixed State: The Most Efficient Purification Protocol}
\author{Ping-Xing Chen$^{1,2}$\thanks{%
E-mail: pxchen@nudt.edu.cn}, Lin-Mei Liang$^{1,2}$, Cheng-Zu Li$^2$ and
Ming-Qiu Huang$^2$}
\address{1. Laboratory of Quantum Communication and Quantum Computation, \\
University of Science and Technology of\\
China, Hefei, 230026, P. R. China \\
2. Department of Applied Physics, National University of\\
Defense Technology,\\
Changsha, 410073, \\
P. R. China. }
\date{\today}
\maketitle

\begin{abstract}
It is well known that any entangled mixed state in $2\otimes 2$ systems can
be purified via infinite copies of the mixed state. But can one distill a
pure maximally entangled state from finite copies of a mixed state in any
bipartite system by local operation and classical communication? This is
more meaningful in practical application. We give a necessary and sufficient
condition of this distillability. This condition can be expressed as: there
exists distillable-subspaces. According to this condition, one can judge
whether a mixed state is distillable or not easily. We also analyze some
properties of distillable-subspaces, and discuss the most efficient
purification protocols. Finally, we discuss the distillable enanglement of
two-quibt system for the case of finite copies.
\end{abstract}


\thispagestyle{empty}

\newpage \pagenumbering{arabic} 

Maximally entangled states have many applications in quantum information,
such as error correcting code\cite{1}, dense coding\cite{2} and teleportation%
\cite{3}, etc. In the laboratory, however, maximally entangled state became
a mixed state easily due to the interaction with environment. This results
in poor application. It involve a basic question: how to distill a pure
entangled state from a mixed state by local operation and classical
communication(LOCC)? Bennett et al\cite{4} proposed a entanglement
purification scheme for a class of Werner states, then Horodecki et al \cite
{5}proved that any inseparable state in $2\otimes 2$ systems(or two-qubit
system) can be distilled into a singlet via this scheme by infinite rounds
of purification protocols, and the necessary condition of distillability\cite
{6} is negative partial transpose(NPT)\cite{7} for any bipartite systems.
This scheme ask only the output states with fidelity $F\rightarrow 1$ under
infinite copies of the purified mixed state. This means that, for one thing,
one can get some states with desirous fidelity from many copies, and get a
near ''perfect''singlet from the infinite number of copies of a mixed state,
for another, one may not get a singlet from some ''distillable'' states
because there are no infinite copies of a mixed state in laboratory. A
natural question is : Can one get a pure entangled state from the finite
number copies of a mixed state? Recently, it has been shown that no
distillation scheme with individual measurement can produce a pure entangled
state from a mixed state of $2\otimes 2$ systems\cite{8,9}. More recently it
was proved that many copies of some mixed state, even if which are almost a
maximally entangled state, also can not produce a pure entangled state\cite
{10}. In this paper, we consider a case of distillability: one can get a
pure maximally entangled state (or a pure entangled state, because pure
entangled states can be transferred reversibly into maximally entangled
states) from the finite number of copies of the mixed states, i.e., ask the
output states with fidelity $F=1$ under finite copies. We prove that the
necessary and sufficient condition of the distillability for any bipartite
systems is that there exists a distillable-subspace(DSS). We analyze some
properties of a DSS, and give an example to demonstrate how to find a DSS.
It is not difficult for one to judge a DSS via this condition. From the
concept of DSS one can get the most efficient purification protocols.
Finally, we discuss the distillable entanglement of mixed states in $%
2\otimes 2$ systems for the case of finite copies, and find only a class of
states are distillable.

A mixed state $\rho _{AB}$ in $N_A\otimes $ $N_B$ system($N_A$ , $N_{B\text{ 
}}$ is the dimension of system A, B, respectively) can be expressed as:

\begin{equation}
\rho _{AB}=\sum_{i=1}^k\lambda _i\left| \Psi _{AB}^i\right\rangle
\left\langle \Psi _{AB}^i\right|
\end{equation}
here $\left\{ \left| \Psi _{AB}^i\right\rangle \right\} $ are pure states
with probability $\left\{ \lambda _i\right\} ,$ respectively, $k$ is the
rank of $\rho _{AB}$. We say $\rho _{AB}$ is $n$-distillable iff from $n$
copies of $\rho _{AB},\rho _{AB}^{\otimes n},$ one can get a pure entangled
state $\left| \Psi \right\rangle $ at least,

\begin{equation}
\left| \Psi \right\rangle =\sum_{i=1}^ma_i\left| e_i\right\rangle _A\left|
f_i\right\rangle _B  \label{1}
\end{equation}
here $\left\{ \left| e_i\right\rangle _A\right\} $ and$\left\{ \left|
f_i\right\rangle _B\right\} $ is a set of orthonormal bases of Hilbert space
of A and B system, respectively, and all $a_i\neq 0,m\geqslant 2.$ If all $%
a_i\neq 0$ ($i=1,...,m$) in Eq(\ref{1}) we say the Schmidt number of pure
state $\left| \Psi \right\rangle $ is $m$

Theorem 1.1: One can distill a pure entangled state $\left| \Psi
\right\rangle $ in Eq(\ref{1}) from $\rho _{AB}^{\otimes 1}$, i.e., $\rho
_{AB}$ is 1-distillable iff: 1), all of $\left| \Psi _{AB}^i\right\rangle s$
include a pure state $\left| \Phi \right\rangle $(e.g., we say the state $%
\frac 1{\sqrt{3}}(\left| 00\right\rangle +\left| 11\right\rangle +\left|
23\right\rangle )$ includes the state $(\left| 00\right\rangle +\left|
11\right\rangle )/\sqrt{2})$

\begin{equation}
\left| \Phi \right\rangle =\sum_{i=1}^mb_i\left| e_i^{^{\prime
}}\right\rangle _A\left| f_i^{^{\prime }}\right\rangle _B  \label{22}
\end{equation}
where all $b_i\neq 0,m\geqslant 2.$ $\left\{ \left| e_i^{^{\prime
}}\right\rangle _A\right\} $ and$\left\{ \left| f_i^{^{\prime
}}\right\rangle _B\right\} $ are another set of orthonormal bases of Hilbert
space of A and B system, respectively; or 2), some $\left| \Psi
_{AB}^i\right\rangle $ include $\left| \Phi \right\rangle $ and the others
which do not include $\left| \Phi \right\rangle $ have not component $%
\left\{ \left| e_i^{^{\prime }}\right\rangle _A\left| f_j^{^{\prime
}}\right\rangle _B\right\} $($i,j=1,...,m$).

Proof: If $\rho _{AB}$ satisfy the condition in theorem1.1, one can first
distill the pure $\left| \Phi \right\rangle $with nonzero probability by
project operation $P_A$ and $P_B:$

\begin{eqnarray}
P_A &=&\sum_{i=1}^m\left| e_i^{^{\prime }}\right\rangle _A\left\langle
e_i^{^{\prime }}\right| _A  \label{3} \\
P_B &=&\sum_{i=1}^m\left| f_i^{^{\prime }}\right\rangle _B\left\langle
f_i^{^{\prime }}\right| _B  \nonumber
\end{eqnarray}
$P_A$ and $P_B$ act on Hilbert space of system A and B, respectively. Then,
transfer the bases $\left| e_i^{^{\prime }}\right\rangle _A\left|
f_i^{^{\prime }}\right\rangle _B$ into $\left| e_i\right\rangle _A\left|
f_i\right\rangle _B$ by local unitary transformation on $\rho _{AB},$ and
get pure $\left| \Phi ^{^{\prime }}\right\rangle $:

\begin{equation}
\left| \Phi ^{^{\prime }}\right\rangle =\sum_{i=1}^mb_i\left|
e_i\right\rangle _A\left| f_i\right\rangle _B  \label{33}
\end{equation}
Finally, one transfer $\left| \Phi ^{^{\prime }}\right\rangle $ into $\left|
\Psi \right\rangle $ by local filter operation\cite{12}. Conversely, suppose
that $\rho _{AB}$ does not satisfy the condition in the theorem 1.1. This
include two cases: 1. there are not $\left| \Phi \right\rangle $ in all pure
states $\left| \Psi _{AB}^i\right\rangle $ under any local unitary
transformation, i.e., $\rho _{AB}$ is separable; 2. Some $\left| \Psi
_{AB}^i\right\rangle $ have component $\left| \Phi \right\rangle $, but the
others which do not include $\left| \Phi \right\rangle $ have ''impure
component'' $\left| e_i^{^{\prime }}\right\rangle _A\left| f_j^{^{\prime
}}\right\rangle _B.$ Obviously, for the first case one cannot distill $%
\left| \Phi \right\rangle $ from $\rho _{AB}.$ For the second case, one must
discard the ''impure component'' to get pure entangled state. To achieve
this, one should distinguish locally\cite{13} the state $\left| \Phi
\right\rangle $ from a state $\left| e_i^{^{\prime }}\right\rangle _A\left|
f_j^{^{\prime }}\right\rangle _B$ without destruction of $\left| \Phi
\right\rangle $. But this is impossible, because both the state $\left| \Phi
\right\rangle $ and the state $\left| e_i^{^{\prime }}\right\rangle _A\left|
f_j^{^{\prime }}\right\rangle _B$ include $\left| e_i^{^{\prime
}}\right\rangle _A$ of system A and $\left| f_j^{^{\prime }}\right\rangle _B$
of system B. One cannot distinguish locally the state $\left| e_i^{^{\prime
}}\right\rangle _A$ ( or $\left| f_j^{^{\prime }}\right\rangle _B$) in state 
$\left| \Phi \right\rangle $ from the state $\left| e_i^{^{\prime
}}\right\rangle _A$ ( or $\left| f_j^{^{\prime }}\right\rangle _B$) in state 
$\left| e_i^{^{\prime }}\right\rangle _A\left| f_j^{^{\prime }}\right\rangle
_B.$ In other words, if one can do so, one can get $\left| \Phi
\right\rangle $ with probability $\lambda $ from a mixed state $\rho
=\lambda |\Phi \rangle \langle \Phi |+(1-\lambda )$ $\left| e_i^{^{\prime
}}\right\rangle _A\left| f_j^{^{\prime }}\right\rangle _B\left\langle
e_i^{^{\prime }}\right| _A$ $\left\langle f_j^{^{\prime }}\right| _B,$ and
the distillable entanglement of $\rho ,E_D(\rho )=$ $\lambda E(|\Phi \rangle
)\geqslant E_F(|\Phi \rangle ),$ here $E(|\Phi \rangle )$ is entanglement of
pure state $\left| \Phi \right\rangle $ , $E_F$ is formation entanglement%
\cite{14}. This inequality cannot hold obviously\cite{11}. So one cannot
discard the ''impure component'' without destruction of $\left| \Phi
\right\rangle .$ Thus we finish the proof of theorem 1.1.

The theorem above is also fit to the case of $\rho _{AB}^{\otimes n}$ , if
we regard $\rho _{AB}^{\otimes n}$ as a state in ($N_A)^{\otimes n}\otimes
(N_B)^{\otimes n}$ systems. It is to say that theorem1.1 can be generalized
into the following theorem:

Theorem 1.2: $\rho _{AB}$ is n-distillable iff all pure states $\left| \Psi
_{AB}^i\right\rangle $ in the pure state decomposition of $\rho
_{AB}^{\otimes n}$ include a pure state $\left| \Phi \right\rangle ,$ or
some $\left| \Psi _{AB}^i\right\rangle $ include $\left| \Phi \right\rangle $
and the others which do not include $\left| \Phi \right\rangle $ have not
component $\left\{ \left| e_i^{^{\prime }}\right\rangle _A\left|
f_j^{^{\prime }}\right\rangle _B\right\} $(i,j=1,...,m). Where $\left\{
\left| e_i^{^{\prime }}\right\rangle _A\right\} $ and$\left\{ \left|
f_i^{^{\prime }}\right\rangle _B\right\} $ are orthonormal vectors of
Hilbert space of $A^{\otimes n}$ and $B^{\otimes n}$ systems, respectively,
and $\left| \Phi \right\rangle =\sum_{i=1}^mb_i\left| e_i^{^{\prime
}}\right\rangle _A\left| f_i^{^{\prime }}\right\rangle _B,$ $b_i\neq
0,m\geqslant 2.$

In essence, theorem 1 show that if one can distill a pure entangled state $%
\left| \Psi \right\rangle $ in equation(\ref{1}) from $\rho _{AB}^{\otimes
n},$ there should exist a subspace $H_{m\otimes m},$ the dimension of which
is $m\otimes m.$ The component of $\rho _{AB}^{\otimes n}$ in this subspace
is a pure state $\left| \Phi \right\rangle $ with same Schmidt number as $%
\left| \Psi \right\rangle $. The distillation protocol is just to project $%
\rho _{AB}^{\otimes n}$ onto this subspace and project out the pure state $%
\left| \Phi \right\rangle $. We define this subspace as distillable-subspace
(DSS).

If $\rho _{AB}^{\otimes n}$ is distillable, $\rho _{AB}^{\otimes n}$ has at
least a DSS $H_{m\otimes m}$ $(m\geqslant 2).$ Because the component of $%
\rho _{AB}^{\otimes n}$ in the DSS $H_{m\otimes m}$ is a pure state $\left|
\Phi \right\rangle ,$ so if we write down the matrix of $\rho _{AB}^{\otimes
n}$ under the product bases $\left\{ \left| e_i^{^{\prime }}\right\rangle
_A\left| f_j^{^{\prime }}\right\rangle _B\right\} $($i,j=1,...,m$) there are 
$m^2-m(m\geqslant 2)$ rows zero elements and $m^2-m(m\geqslant 2)$ columns
zero elements in the matrix of $\rho _{AB}^{\otimes n}$. The rank of $\rho
_{AB}^{\otimes n}$ is at almost $N_A^n.N_B^n-m^2+1$, for the rank of a DSS $%
H_{m\otimes m}$ is one. Thus we can get the following conclusion:

Theorem 2: If $\rho _{AB}$ is $n$-distillable, the rank of $\rho
_{AB}^{\otimes n}$ is at almost $N_A^n.N_B^n-m^2+1(m\geqslant 2),$ and $\rho
_{AB}^{\otimes n}$ can be expressed as a form with $(m^2-m)(m\geqslant 2)$
rows elements and $(m^2-m)(m\geqslant 2)$ columns elements being zero.

Theorem 2 imply that all mixed states $\rho _{AB}$ in $2\otimes 2$ systems
cannot be distilled by individual copy, as is acclaimed before\cite{9,10}.
Because the dimension of Hilbert space for any DSS is equal to or more than
4, if one can distill a pure entangled state from $2\otimes 2$ systems, then
the whole space of the $2\otimes 2$ systems will be a DSS and $\rho _{AB}$
be a pure state. In fact, there are some mixed states in $2\otimes 2$
systems, even if they are entangled state one cannot distill a pure
entangled state from $\rho _{AB}^{\otimes n}$ if $n$ is finite\cite{10}.

Obviously, the DSS have the following properties.

1. The component of $\rho _{AB}^{\otimes n}$ in a DSS is a pure entangled
state.

2. From the example in the following, it can be shown that $\rho
_{AB}^{\otimes n}$ may has many DSS, and a few DSS may be combined into a
DSS.

3. Any LOCC cannot produce a new extra DSS without the destruction of
existing DSS owing to entanglement non-increasing under LOCC.

Now, we give an example to demonstrate how to judge whether a mixed state $%
\rho _{AB}($or $\rho _{AB}^{\otimes n})$ has a DSS or not. We have a mixed
state in $2\otimes 2$ systems:

\begin{equation}
\rho _{AB}=\left[ 
\begin{array}{llll}
\frac 14 & 0 & 0 & \frac 14 \\ 
0 & \frac 12 & 0 & 0 \\ 
0 & 0 & 0 & 0 \\ 
\frac 14 & 0 & 0 & \frac 14
\end{array}
\right]  \label{5}
\end{equation}
One can find the DSS of $\rho _{AB}^{\otimes 2}$ in following steps:

1. Calculate the eigenvectors $\left| \Psi _{AB}^i\right\rangle $ and the
nonzero eigenvalues $\lambda _i$ of $\rho _{AB}$. In this case, $\left| \Psi
_{AB}^1\right\rangle =(\left| \uparrow \right\rangle \left| \uparrow
\right\rangle +\left| \downarrow \right\rangle \left| \downarrow
\right\rangle )/\sqrt{2},\left| \Psi _{AB}^2\right\rangle =\left| \uparrow
\right\rangle \left| \downarrow \right\rangle ;\lambda _1=\lambda _2=1/2.$

2. Write all pure states of $\rho _{AB}^{\otimes 2}.$

\begin{eqnarray}
\lambda _1^2 &:&(\left| 0\right\rangle \left| 0\right\rangle +\left|
1\right\rangle \left| 1\right\rangle +\left| 2\right\rangle \left|
2\right\rangle +\left| 3\right\rangle \left| 3\right\rangle )/2;\ \ \ \
\lambda _1\lambda _2:(\left| 0\right\rangle \left| 2\right\rangle +\left|
1\right\rangle \left| 3\right\rangle )/\sqrt{2}  \label{6} \\
\ \lambda _1\lambda _2 &:&(\left| 0\right\rangle \left| 1\right\rangle
+\left| 2\right\rangle \left| 3\right\rangle )/\sqrt{2};\ \ \ \ \ \ \ \ \ \
\ \ \ \ \ \ \ \ \ \ \ \ \lambda _2^2=\left| 0\right\rangle \left|
3\right\rangle  \nonumber
\end{eqnarray}
where $\left| 0\right\rangle =\left| \uparrow \uparrow \right\rangle ,\left|
1\right\rangle =\left| \uparrow \downarrow \right\rangle ,\left|
2\right\rangle =\left| \downarrow \uparrow \right\rangle ,\left|
3\right\rangle =\left| \downarrow \downarrow \right\rangle ,$ are similar to
binary form.

3. Find the DSS from all pure states above. In this case we find easily a
DSS with probability $1/2\lambda _1^2$, in which the state $(\left|
1\right\rangle \left| 1\right\rangle +\left| 2\right\rangle \left|
2\right\rangle )/\sqrt{2}$ can be distilled. We can write down the pure
decomposition of $\rho _{AB}^{\otimes n}$ with the method in step 2 and can
find the all DSS of $\rho _{AB}^{\otimes n}$ by symmetry(see Fig.1). For
example, the DSS of $\rho _{AB}^{\otimes 3}$ are(we represent the DSS with
corresponding distillable pure state from this DSS): $(\left| 1\right\rangle
\left| 1\right\rangle +\left| 2\right\rangle \left| 2\right\rangle )/\sqrt{2}
$ with probability $1/4\lambda _1^3$; $(\left| 1\right\rangle \left|
5\right\rangle +\left| 2\right\rangle \left| 6\right\rangle )/\sqrt{2}$ with
probability $1/2\lambda _1^2\lambda _2;$ $(\left| 5\right\rangle \left|
5\right\rangle +\left| 6\right\rangle \left| 6\right\rangle )/\sqrt{2}$ with
probability $1/4\lambda _1^3;$ $(\left| 3\right\rangle \left| 3\right\rangle
+\left| 4\right\rangle \left| 4\right\rangle )/\sqrt{2}$ with probability $%
1/4\lambda _1^3$. Or, the four lower dimension DSS above can be combined
into two higher dimension DSS: $(\left| 1\right\rangle \left| 1\right\rangle
+\left| 2\right\rangle \left| 2\right\rangle +\left| 4\right\rangle \left|
4\right\rangle )/\sqrt{3}$ with probability $3/8\lambda _1^3$ and $(\left|
3\right\rangle \left| 3\right\rangle +\left| 5\right\rangle \left|
5\right\rangle +\left| 6\right\rangle \left| 6\right\rangle )/\sqrt{3}$ with
probability $3/8\lambda _1^3$.

Now, we discuss the most efficient protocols for the entanglement purifying$%
. $ For the case of finite copies and asking the output with unite fidelity,
from the proof of theorem1 and the conception of DSS we can get that the
most efficient protocols for entanglement distillation from $\rho
_{AB}^{\otimes n}$ is to project out, and not destroy, every independent DSS
of $\rho _{AB}^{\otimes n}$ with corresponding probability, this is because
LOCC cannot produce a extra DSS, but may destroy the DSS. For example, to
get the two DSS of $\rho _{AB}^{\otimes 3}$ in the above example, $(\left|
1\right\rangle \left| 1\right\rangle +\left| 2\right\rangle \left|
2\right\rangle +\left| 4\right\rangle \left| 4\right\rangle )/\sqrt{3}$ with 
$3/8\lambda _1^3$ probability and $(\left| 3\right\rangle \left|
3\right\rangle +\left| 5\right\rangle \left| 5\right\rangle +\left|
6\right\rangle \left| 6\right\rangle )/\sqrt{3}$ with $3/8\lambda _1^3$
probability, after local unitary transformations one may use three local
project operation :

\begin{eqnarray}
P_1 &=&\left| 1\right\rangle \left\langle 1\right| +\left| 2\right\rangle
\left\langle 2\right| +\left| 4\right\rangle \left\langle 4\right|  \label{8}
\\
P_2 &=&\left| 3\right\rangle \left\langle 3\right| +\left| 5\right\rangle
\left\langle 5\right| +\left| 6\right\rangle \left\langle 6\right|  \nonumber
\\
P_3 &=&\left| 0\right\rangle \left\langle 0\right| +\left| 7\right\rangle
\left\langle 7\right|  \nonumber
\end{eqnarray}
onto A system, then A system sends the output result to B system. After
receiving the information from A, B system use the same project operation.
Consequently, one get the two DSS with same probability $3/8\lambda _1^3,$
and the other state with probability ($1-3/4\lambda _1^3).$ As Ref\cite{1}
mentioned, all distillation protocols involve one-way or two-way
communication. The efficiency of one-way is higher than two-way. Obviously,
the distillation protocol here involve only one-way communication. According
to the most efficient protocol one can calculate the distillable
entanglement for the case of the finite copies.

Similarly, for the case of infinite copies, the most efficient purifying
protocols of $\rho _{AB}$ is to let $n$-copy of $\rho _{AB}$ $(n\rightarrow
\infty )$ spans a bigger Hilbert space, then project out the desirous
subspace of $\rho _{AB}^{\otimes n}$ by PostSelection operation\cite{11}.
The component of $\rho _{AB}^{\otimes n}$ in the desirous subspace tends to
a pure entangled state when $n\rightarrow \infty .$ The whole purifying
process ask only a round of purifying protocols. In the purifying scheme in
Ref\cite{4}, every round of purifying is also to keep the desirous subspace,
the bases of which is $\left| 1\right\rangle \left| 1\right\rangle ,\left|
1\right\rangle \left| 2\right\rangle ,\left| 2\right\rangle \left|
1\right\rangle ,\left| 2\right\rangle \left| 2\right\rangle $(these bases
have same definition as in Eq(\ref{6})), but the whole purifying process
needs many rounds of purifying protocols which result in the unavoidable
loss of distillable entanglement , so this scheme is not the most efficient
one. How to calculate the distillable entanglement for infinite-copy case is
beyond the realm of this paper.

Now, we will discuss the distillable entanglement $E_D$ of mixed states in $%
2\otimes 2$ systems for the case of finite copies. Here we will show that
only a class of states are distillable.

First, a necessary condition of distillability from $\rho _{AB}^{\otimes n}$
is that $\rho _{AB}^{\otimes n}$ is not a quasi-separable state(QSS)\cite{10}%
. We say a state $\rho $ is a QSS iff one or many {\it new-state} of $\rho $
is separable. Any mixed state $\rho $ has infinite sets of pure state
decompositions.

\begin{equation}
\rho =\sum_{i=1}p_i\left| \Psi _i\right\rangle \left\langle \Psi _i\right|
\label{77}
\end{equation}
For every decomposition, if one lets the pure state $\left| \Psi
_i\right\rangle $ unchanged but change the probability $p_i$ of pure state $%
\left| \Psi _i\right\rangle $ in the real numbers realm (0,1), we say one
gets a {\it new-state} of $\rho $\cite{10}.

Second, a mixed state $\rho $ can be decomposed into\cite{14} 
\begin{equation}
\rho =\sum_{i=1}^l\left| x_i\right\rangle \left\langle x_i\right|
=\sum_{i=1}^m\left| z_i\right\rangle \left\langle z_i\right| ,  \label{88}
\end{equation}
where $\left| x_i\right\rangle ,$ unnormalized, is a complete set of
orthogonal eigenvectors corresponding to the nonzero eigenvalues of $\rho $,
and $\left\langle x_i\right| x_i\rangle $ is equal to the its nonzero
eigenvalues. For a state of $2\otimes 2$ systems, there exist a set of
decomposition $\left| z_i\right\rangle $ of $\rho $ noted by 
\begin{equation}
\left| z_i\right\rangle =\sum_{j=1}^lu_{ij}\left| x_j\right\rangle ,\qquad
i=1,2,\cdots ,m  \label{9}
\end{equation}
where $\left| z_i\right\rangle $ is not necessarily orthogonal, the columns
of transformation $u_{k\times l}$ are orthonormal vectors, and

\begin{equation}
\left\langle z_i\right| \left. \widetilde{z}_j\right\rangle =\lambda
_i^{^{\prime }}\delta _{ij},  \label{10}
\end{equation}
where, $\left| \widetilde{z}_i\right\rangle =\sigma _y\otimes \sigma
_y\left| z_i^{*}\right\rangle $, and $\lambda _1^{^{\prime }}>\lambda
_2^{^{\prime }}>\lambda _3^{^{\prime }}>$ $\lambda _4^{^{\prime }}\geqslant
0 $. Let us suppose $\lambda _1^{^{\prime }}-\lambda _2^{^{\prime }}-\lambda
_3^{^{\prime }}-\lambda _4^{^{\prime }}>0,$ namely $\rho $ is inseparable%
\cite{14}. If not all $\lambda _i^{^{\prime }}$($i=2,3,4$) being zero, one
can transfer the state $\rho $ into a separable state by decreasing the
probability appearing $\left| z_1\right\rangle $, so the state $\rho $ is a
QSS and $\rho ^{\otimes n}$ is also a QSS. Thus $\rho $ cannot be distilled,
i.e., $E_D(\rho )=0.$

Third, a mixed state $\rho $ of $2\otimes 2$ systems can be expressed as\cite
{8}:

\begin{equation}
\rho =\lambda \left| \Psi \right\rangle \left\langle \Psi \right|
+(1-\lambda )\rho _{sep}  \label{11}
\end{equation}
$\rho _{sep}$ is a separable state. $\rho $ surely includes a mixed state $%
\rho _1$:

\begin{equation}
\rho _1=\lambda _1\left| \Psi \right\rangle \left\langle \Psi \right|
+(1-\lambda _1)\left| \Phi \right\rangle \left\langle \Phi \right|
\label{12}
\end{equation}
where

\begin{equation}
\left| \Psi \right\rangle =\sin \theta \left| 0\right\rangle _A\left|
0\right\rangle _B+\cos \theta \left| 1\right\rangle _A\left| 1\right\rangle
_B  \label{13}
\end{equation}

\begin{equation}
\left| \Phi \right\rangle =(a_1\left| 0\right\rangle _A+a_2\left|
1\right\rangle _A)\otimes (b_1\left| 0\right\rangle _B+b_2\left|
1\right\rangle _B)  \label{14}
\end{equation}
$\left| 0\right\rangle _i$ and $\left| 1\right\rangle _i$ $(i=A$ or $B)$ are
any orthogonal bases of Alice's or Bob's system, respectively. It is to say $%
\rho $ is a mixture of $\rho _1$ and a separable state. So if $\rho $ is
distillable, then $\rho _1$ is also distillable and $\rho _1$ should be not
a QSS. If $a_1,a_2,b_1,b_2$ are all nonzero, the matrix of $\rho _1$ and $%
\rho _1^{\otimes n}$ have not zero diagonal elements under any product bases
(i.e., the bases vectors are not separable states). From the theorem 2 one
can follow that $E_D(\rho _1)=0.$ If $\left| \Phi \right\rangle =(a_1\left|
0\right\rangle _A+a_2\left| 1\right\rangle _A)\otimes \left| 0\right\rangle
_B$ or $\left| \Phi \right\rangle =\left| 0\right\rangle _A\otimes
(b_1\left| 0\right\rangle _B+b_2\left| 1\right\rangle _B),$ $\rho _1$ have
two nonzero $\lambda _i^{^{\prime }}$(see Eq(\ref{10})). Thus $\rho _1$ is a
QSS, and $E_D(\rho _1)=0.$ So only when $\left| \Phi \right\rangle =\left|
0\right\rangle _A\left| 1\right\rangle _B$ or $\left| \Phi \right\rangle
=\left| 1\right\rangle _A\left| 0\right\rangle _B,$ $E_D(\rho _1)\neq 0.$
But if $\rho =\lambda _1\left| \Psi \right\rangle \left\langle \Psi \right|
+\lambda _2\left| 0\right\rangle _A\left| 1\right\rangle _B\left\langle
0\right| _A\left\langle 1\right| _B+\lambda _3\left| 1\right\rangle _A\left|
0\right\rangle _B\left\langle 1\right| _A\left\langle 0\right| _B,$ $%
E_D(\rho )\neq 0$ for the matrix of $\rho $ and $\rho ^{\otimes n}$ have not
zero diagonal elements under any product bases. So only $\rho $ with
following forms have nonzero distillable entanglement: 
\begin{equation}
\rho =\lambda _1\left| \Psi ^{\prime }\right\rangle \left\langle \Psi
^{\prime }\right| +\lambda _2\left| \Phi ^{\prime }\right\rangle
\left\langle \Phi ^{\prime }\right|  \label{15}
\end{equation}
or 
\begin{equation}
\rho =\lambda _1\left| \Psi ^{^{^{\prime \prime }}}\right\rangle
\left\langle \Psi ^{^{\prime \prime }}\right| +\lambda _2\left| \Phi
^{^{\prime \prime }}\right\rangle \left\langle \Phi ^{\prime \prime }\right|
\label{16}
\end{equation}
where $\lambda _1+\lambda _2=1,$ $\left| \Psi ^{\prime }\right\rangle =\sin
\theta \left| 0\right\rangle _A\left| 0\right\rangle _B+\cos \theta \left|
1\right\rangle _A\left| 1\right\rangle _B,\left| \Phi ^{\prime
}\right\rangle =\left| 01\right\rangle $ or $\left| 10\right\rangle ;$ $%
\left| \Psi ^{\prime \prime }\right\rangle =\sin \theta \left|
0\right\rangle _A\left| 1\right\rangle _B+\cos \theta \left| 1\right\rangle
_A\left| 0\right\rangle _B,\left| \Phi ^{\prime }\right\rangle =\left|
00\right\rangle $ or $\left| 11\right\rangle .$

Now, we would like to discuss the relations between the DSS which emerge in $%
n$ copies of uncorrelated pairs and decoherence-free-subspaces which are due
to some collective noise. Suppose that each of many pure singlet pairs
became the same mixed states owing to interaction with environment. Although
each one became a mixed state, there may be some subspaces with pure
entangled states in the whole Hilbert space of all pairs. These subspaces
are decoherence-free. So in this sense, DSS are decoherence-free-spaces, and
somehow allow for perfect error correction.

In summary, one can distill a pure entangled state iff there exists a
distillable-subspace in which the component of $\rho _{AB}$ is a pure state
with Schmidt number $m\geqslant 2$. If there exists the
distillable-subspace, one can get a pure entangled state by the project
operation. It is not different for one to find a distillable-subspace of $%
\rho _{AB},$ and to get all distillable-subspace of $\rho _{AB}^{\otimes n}$
with symmetry. The most efficient distillation protocols(include the case of
both infinite copies and finite copies) is to keep the desirous subspace and
discard the other subspace by project operation. For the case of finite
copies of mixed states in $2\otimes 2$ system, only a class of state are
distillable.

\acknowledgments  We thank professor C.H.Bennett for his valuable and
extensive comments and suggestion, and professor Guangcan Guo for his help
to this work.

\begin{center}
\begin{minipage}{120mm}
{\sf Fig. 1.} \small{ The pure-state decomposition of $\rho _{AB}^{\otimes n}.$ Each pure
state is expressed by some binary numbers. According to the symmetry, one
can write down all pure state of $\rho _{AB}^{\otimes n}$. One can also find
the independent DSS of $\rho _{AB}^{\otimes n}$. }
\end{minipage}
\begin{figure}[htbp]
\end{figure}
\end{center}

\end{document}